\documentstyle[eqsecnum,aps,epsfig]{revtex}
\begin{document}
\draft
%\wideabs{
\title{
Computing radiation from Kerr black holes:\\
Generalization of the Sasaki-Nakamura equation
}
\author{Scott A.\ Hughes}
\address{
Theoretical Astrophysics, California Institute of Technology,
Pasadena, CA 91125}
\maketitle
\begin{abstract}
As shown by Teukolsky, the master equation governing the propagation
of weak radiation in a black hole spacetime can be separated into four
ordinary differential equations, one for each spacetime coordinate.
(``Weak'' means the radiation's amplitude is small enough that its own
gravitation may be neglected.)  Unfortunately, it is difficult to
accurately compute solutions to the separated radial equation (the
Teukolsky equation), particularly in a numerical implementation.  The
fundamental reason for this is that the Teukolsky equation's
potentials are long ranged.  For non-spinning black holes, one can get
around this difficulty by applying transformations which relate the
Teukolsky solution to solutions of the Regge-Wheeler equation, which
has a short-ranged potential.  A particularly attractive
generalization of this approach to spinning black holes for
gravitational radiation (spin weight $s = -2$) was given by Sasaki and
Nakamura.  In this paper, I generalize Sasaki and Nakamura's results
to encompass radiation fields of arbitrary integer spin weight, and
give results directly applicable to scalar ($s = 0$) and
electromagnetic ($s = -1$) radiation.  These results may be of
interest for studies of astrophysical radiation processes near black
holes, and of programs to compute radiation reaction forces in curved
spacetime.
\end{abstract}
%}

\section{Introduction}
\label{sec:intro}

In 1973, Teukolsky {\cite{teuk73}} derived a single partial
differential equation describing the evolution of perturbations to
rotating (Kerr) black holes.  This master equation gives the
linearized evolution of fields that arise from a perturbing source of
stress energy --- the charge and current densities associated with the
perturbation --- to the (vacuum) black hole background.  The solutions
of the homogeneous version of this equation describe the propagation
of radiation in black hole spacetimes.  Thus, a common use of this
formalism is to study the radiation emitted by matter in the
environment of a black hole.  In some cases, one can use such an
analysis to study back reaction, determining how the perturbing source
evolves as radiation carries away energy and angular momentum.  A
beautiful feature of the master equation is that it describes
radiation fields of arbitrary spin weight $s$.  It has been used
extensively to study scalar ($s = 0$), electromagnetic ($s = \pm 1$),
and gravitational ($s = \pm 2$) radiation in Kerr spacetimes.

The master equation is often\footnote{Separation is not always used.
There is also a body of work that uses the master equation to evolve
initial data.  This approach has been extensively used to study the
endpoint of binary black hole collisions; see Ref.\ {\cite{pullin}}
and references therein.} solved by introducing a multipolar
decomposition of the radiation field.  The solution separates into
functions of the Boyer-Lindquist coordinates:
\begin{equation}
{_s}\Psi = \sum_{l,m,\omega} R_{lm\omega}(r) {_s}S_{lm}^{a\omega}(\theta)
e^{im\phi} e^{-i\omega t}\;;
\label{eq:multipolar}
\end{equation}
each function is governed by an ordinary differential equation.  (The
precise meaning of ${_s}\Psi$ is described in Sec.\
{\ref{sec:teuk_gen}}.)  The $t$ and $\phi$ dependences are trivial.
The $\theta$ dependence is more involved, but can be evaluated in a
straightforward matter.  The functions ${_s}S_{lm}^{a\omega}$ are
spin-weighted spheroidal harmonics, which are generalizations of
spin-weighted spherical harmonics to a spheroidal geometry.  The
spin-weighted spherical harmonics in turn are generalizations of
spherical harmonics that encode the rotation properties of spin $s$
fields; see Refs.\ {\cite{spherical1,spherical2}} for further
discussion.  A detailed algorithm for computing
${_s}S^{a\omega}_{lm}(\theta)$ is given in Ref.\
{\cite{sah_kerr_gwI}}.  For the purposes of this paper, the $\theta$
dependence is considered known.

The radial dependence, $R_{lm\omega}(r) \equiv R(r)$, on the other
hand, can be rather difficult to calculate in practice, particularly
in a numerical computation.  The fundamental reason for this
difficulty is the nature of the equation that governs $R(r)$: this
equation (the Teukolsky equation) has a long-ranged potential.  In
source-free form, it can be written
\begin{equation}
{d^2R\over dr^{*2}} + F_T(r) {dR\over dr^*} + \left[\omega^2
- U_T(r)\right] R = 0\;,
\label{eq:teuk_rstar}
\end{equation}
where $\omega$ is the frequency of the radiation mode and
\begin{equation}
r^* = r + {2 M r_+\over r_+ - r_-}\ln{r - r_+\over 2M}
- {2 M r_-\over r_+ - r_-}\ln{r - r_-\over 2M}
\label{eq:kerr_tortoise}
\end{equation}
is the Kerr ``tortoise coordinate''.  The potentials $U_T(r)$ and
$F_T(r)$ are rather complicated; they encode the most interesting
features of wave propagation in black hole spacetimes, such as scatter
from spacetime curvature\footnote{As shown by Leonard and Poisson
{\cite{leonard_poisson}}, the phenomenon of {\it tails} (delayed
propagation due to scatter from spacetime curvature) is to leading
order independent of $s$, and is encoded in the logarithmic behavior
of $r^*$, not the potentials.} and superradiant scattering (radiation
whose scattered amplitude exceeds the ingoing amplitude due to
extraction of energy from the black hole's spin).  For large $r$,
\begin{eqnarray}
F_T(r) &=& {2(1 + s)\over r} - {2M(2 + s)\over r^2} + O(1/r^3)\;,
\nonumber\\
U_T(r) &=& -{4is\omega\over r} + {\lambda + 2am\omega +
8IMs\omega\over r^2} + O(1/r^3)\;.
\label{eq:teuk_potentials_limit}
\end{eqnarray}
[The quantity $\lambda$ is related to the eigenvalues of the $\theta$
dependence; see Ref.\ {\cite{sah_kerr_gwI}} for details.]  For large
$r$, $F_T(r)$ and $U_T(r)$ fall off only as $1/r$ --- they are
long-ranged, like the Coulomb potential.  The solution of Eq.\
(\ref{eq:teuk_rstar}) for large $r$ is {\cite{teukpress}}
\begin{equation}
R = C_1 {e^{-i\omega r^*}\over r} + C_2 {e^{i\omega r^*}\over r^{2s+1}}\;.
\label{eq:teuk_soln_inf}
\end{equation}
The complex constants $C_1$ and $C_2$ are determined by boundary
conditions.  This asymptotic solution illustrates the difficulty in
solving the Teukolsky equation: the coefficient of $e^{i \omega r^*}$
differs from the coefficient of $e^{-i \omega r^*}$ by $r^{-2s}$.  For
negative $s$, this becomes extremely large --- large enough that the
ingoing $e^{-i\omega r^*}$ piece will eventually be entirely lost in
any numerical computation due to round-off error.  Hence, for negative
$s$, it is nearly impossible to set proper boundary conditions on the
solution's phase at large $r$.  Similarly, for positive $s$ it is
difficult to set boundary conditions near the event
horizon\footnote{One can expand the potentials near the horizon and
see that they die away slowly as $r^* \to -\infty$ (which corresponds
to $r \to r_+$).  The actual form is somewhat messy, and is not given
here explicitly.}.  For $r$ very close to $r_+ = M +\sqrt{M^2 - a^2}$
(the location of the event horizon in Boyer-Lindquist coordinates),
the solution is {\cite{teukpress}}
\begin{equation}
R = C_3 \Delta^{-s}e^{-i p r^*} + C_4 e^{i p r^*}\;.
\label{eq:teuk_soln_horiz}
\end{equation}
I have introduced $p = \omega - m\omega_+$, where $\omega_+ = a/2 M
r_+$ is the angular velocity at which observers at the horizon are
seen to rotate.  Because the Boyer-Lindquist coordinates $t$ and
$\phi$ become twisted and entangled near the horizon, $p$ describes
the frequency of wave modes in that region.  The factor $\Delta = r^2
- 2 M r + a^2$ goes to zero at the event horizon.  Hence, for positive
$s$, the ingoing solution swamps the outgoing solution as one
approaches the horizon.  {\it Whether one chooses positive or negative
$s$, there exists a domain in which one cannot accurately compute
numerical solutions by directly integrating the homogeneous Teukolsky
equation.}

Various approaches have been discussed to circumvent this difficulty.
One of the first was introduced by Teukolsky and Press
{\cite{teukpress,pressteuk}}.  Their approach used the fact that, for
a given $|s|$, the solutions to Eq.\ (\ref{eq:teuk_rstar}) for $s =
+|s|$ and $s = -|s|$ are physically equivalent: there exist rules to
take the positive $s$ solution to the negative $s$ solution, and vice
versa.  Thus, one can initially choose $s$ so that the solution is
well-behaved in the initial $r$ domain, and then ``switch horses'' and
integrate with the other sign of $s$ as the integration approaches the
other asymptotic domain.

A somewhat more elegant way to compute $R$ was developed by
Chandrasekhar {\cite{chandra_transform}}.  As already noted, the poor
behavior of the solutions (\ref{eq:teuk_soln_inf}) and
(\ref{eq:teuk_soln_horiz}) is due to the long-ranged nature of the
Teukolsky equation's potentials.  Rather than try to work with an
equation that is simply not well-behaved, one should find
transformations which relate the Teukolsky solution $R$ to the
solution $X$ of some equation whose potentials are short-ranged.  For
example, when the black hole spin is zero, black hole perturbations
can be described using the generalized Regge-Wheeler equation
{\cite{leaver85}}:
\begin{equation}
\left[{d^2\over dr^{*2}} + \omega^2 - V_{RW}(r,s)\right]X = 0\;,
\label{eq:reggewheeler}
\end{equation}
where
\begin{equation}
V_{RW}(r,s) = f\left[{l(l+1)\over r^2} -{2(s^2 - 1)M\over
r^3}\right].
\label{eq:RWpot}
\end{equation}
(Here, $f = 1 - 2 M / r$.)  This potential dies away faster than
$1/r$, and so is short ranged.  Chandrasekhar showed (for specific
choices of $s$) that solutions to Eq.\ (\ref{eq:reggewheeler}) and
solutions to Eq.\ (\ref{eq:teuk_rstar}) (with $a = 0$) are related by
simple rules.  (Below I generalize these rules to any value of $s$.)
This is extremely useful for numerical work: one can integrate Eq.\
(\ref{eq:reggewheeler}) to accurately compute $X$ and then transform
to $R$.  Note that in Ref.\ {\cite{pressteuk}} Press and Teukolsky
introduced a transformation that, in essence, transformed to a
function governed by an equation with a better behaved potential.
They did not, however, discuss the nature of the transformation in
terms of the rangedness of the potentials; Chandrasekhar appears to
have been the first to systematically approach this problem with the
viewpoint that the long-ranged potential was the key issue.  Note also
that Chandrasekhar's notation {\cite{chandra_transform}} is rather
different from that used here; I use a notation similar to that used
in {\cite{poisson93}}.  The transformation given in Refs.\
{\cite{chandra_transform}} and {\cite{poisson93}} is for $s = -2$; a
rule for $s = -1$ is given in {\cite{leonard_poisson}}.

For spinning black holes, perhaps the most elegant generalization of
Chandrasekhar's approach was given by Sasaki and Nakamura
{\cite{sasaknak}}.  They derive a transformation rule which relates
$R$ for any physical spin $a$ to the solution $X$ of an equation whose
potentials are short ranged.  The transformation and short-ranged
potentials are designed such that if $a = 0$, the potentials reduce to
the Regge-Wheeler potentials.  This approach is very natural in the
sense that its solutions are monotonic with respect to spin, ranging
from the Schwarzschild value $a = 0$ to the extreme Kerr limit $a =
M$.  Chandrasekhar and Detweiler also investigated several
transformations relating $R$ to a short-ranged solution $X$
{\cite{chandra_det76,det77,chandra_cent,chandra_mtbh}}.  In some (but
not all) cases, these transformations are governed by equations which
reduce to the Regge-Wheeler equation in the $a = 0$ limit; however,
the equations themselves are often not as ``nice'' to work with.  For
example, the equation for $X$ is sometimes given in terms of a
frequency dependent variable ${\hat r}^*(\omega)$ (see Ref.\
{\cite{chandra_cent}}) which is different from the ``usual'' tortoise
coordinate $r^*$ [cf.\ Eq.\ (\ref{eq:kerr_tortoise})].  This
coordinate can be doubly valued and mask features such as superradiant
scattering.  Also, the potentials of each of their equations are
pathological for some set of frequencies {\cite{det_pc}}.  By using a
set of multiple perturbation equations and transformation laws, one
can always find a non-pathological tool for any given frequency.  But,
there is no single rule that works for all frequencies.  (These
difficulties do not mean that Chandrasekhar and Detweiler's approaches
are not useful.  Campanelli and Lousto {\cite{camp_lousto}} used rules
very similar to Chandrasekhar and Detweiler's in order to show that
solutions to the Teukolsky equation are well-behaved even for sources
that extend to infinity.)

Sasaki and Nakamura's work is restricted to the choice $s = -2$.  This
is an appropriate choice for studies of gravitational perturbations,
and so the Sasaki-Nakamura equation has been extensively used in
studies of gravitational-wave generation and gravitational radiation
reaction
{\cite{sah_kerr_gwI,sasaknak,kojima_nakamura,shibata94,msstt97,kennefick}}.
Other values of $s$ are interesting as well.  For example, $s = \pm 1$
corresponds to electromagnetic radiation.  The propagation and scatter
of electromagnetic waves in black hole spacetimes is of great
astrophysical interest.  Also, much work is currently being directed
toward understanding how one calculates self forces and radiation
reaction forces in curved spacetimes
{\cite{mst97,quinn_wald,and_flan_ott,wiseman,ori,barack_ori,burko1,burko2}}.
Implementations of the general formalism (Refs.\
{\cite{mst97,quinn_wald}}) to date have been restricted to scalar ($s
= 0$) or electromagnetic fields.  They have also been restricted to
spherically symmetric spacetimes.  Tools for effective calculation of
radiation fields in Kerr spacetimes will make it possible to extend
these calculations to more realistic spinning black holes.

In this paper, I generalize the Sasaki-Nakamura equation to arbitrary
integer spin weight $s$.  The generalized Sasaki-Nakamura (GSN)
potentials are given for any $s$, but in terms of two unknown
functions, $\alpha(r)$ and $\beta(r)$.  These functions are fixed by
requiring that the transformation which relates the Teukolsky solution
to the GSN solution reduces, in the Schwarzschild limit, to the
transformation between the Teukolsky solution and the Regge-Wheeler
solution.  They also must be chosen so that the potentials they
generate are of short range.

There is a great deal of freedom in how one chooses $\alpha$ and
$\beta$: for each value of $s$, there are an infinite number of
functions which lead to a transformation with the correct limiting
value and that produces short-ranged potentials.  It is thus most
practical to develop $\alpha$ and $\beta$ on a case by case basis,
rather than trying to develop generic formulas.  Given the interest in
scalar and electromagnetic radiation, I provide examples of $\alpha$
and $\beta$ for $s = 0$ and $s = -1$.

Throughout this paper, a prime denotes $d/dr$, where $r$ is either the
Boyer-Lindquist or the Schwarzschild coordinate (which coordinate
should be clear from context).  An overbar denotes complex
conjugation.  The function $\Delta = r^2 - 2 M r + a^2$, and $f = 1 -
2 M / r$.  Thus, for Schwarzschild holes, $\Delta = r^2 f$.  The
function $R$ will always denote the solution to the homogeneous
Teukolsky equation; $X$ will always refer to the solution of the
equation with short-ranged potential.  Section {\ref{sec:teuk_gen}
reviews important aspects of the Teukolsky equation and its solutions,
particularly how one can compute the solution given an appropriate
source term and solutions to the homogeneous equation, and how those
solutions are related to physical radiation fields.  In Sec.\
{\ref{sec:schwarzschild}, I review the transformation rules for
Schwarzschild black holes.  These rules, which are particularly
simple, will be used as guidelines for constructing transformations
appropriate to radiation in Kerr spacetimes.  Finally, in Sec.\
{\ref{sec:kerr}} I construct the short-ranged equation for Kerr black
holes and provide a recipe for specifying the transformation rule for
a radiation field of arbitrary spin weight.  I apply this recipe in
Secs.\ {\ref{sec:scalar}} and {\ref{sec:electromagnetic}} to scalar
and electromagnetic radiation fields, respectively.  The resultant
transformation rules and equations should form a useful basis for
further studies of radiation in Kerr spacetimes.  Some concluding
discussion is given in Sec.\ {\ref{sec:conclusion}}.

\section{Some properties of the Teukolsky equation and its solutions}
\label{sec:teuk_gen}

As background for the calculations in this paper, I review in this
section the most important properties of the Teukolsky equation
and its solutions.

As discussed in the Introduction, Teukolsky {\cite{teuk73}} showed
that one can separate the wave equation for a field ${_s}\Psi$ of spin
weight $s$ radiation propagating on a Kerr black-hole background using
the multipolar decomposition given in Eq.\ (\ref{eq:multipolar}).  The
$\phi$ and $t$ dependence is trivial, and the $\theta$ dependence is
straightforwardly dealt with.  The $r$ dependence, on the other hand,
can cause problems.

The radial function $R(r)$ is a solution to the Teukolsky equation,
Eq.\ (\ref{eq:teuk_rstar}).  Here I write the Teukolsky equation with
its source term and in terms of derivatives with respect to $r$ rather
than $r^*$:
\begin{equation}
\Delta^{-s}\left(\Delta^{s+1} R'\right)' - V_T(r) R =
-{\cal T}(r)\;.
\label{eq:teuk_eqn_inhom}
\end{equation}
This is the way the Teukolsky equation usually appears in the
literature.  The potential $V_T(r)$ is
\begin{equation}
V_T(r) = \lambda  - 4 i s \omega r -
{{K(r)^2 - 2 i s (r - M)K(r)}\over\Delta}\;;
\label{eq:teuk_pot}
\end{equation}
the quantity $\lambda = {\cal E}_{lm} - 2 a m \omega + a^2\omega^2 -
s(s+1)$, where ${\cal E}_{lm}$ is the eigenvalue of the spheroidal
harmonic [see {\cite{sah_kerr_gwI}}; in the Schwarzschild limit,
${\cal E}_{lm} = l(l+1)$].  The function $K(r) = (r^2 + a^2)\omega -
ma$.

The source term ${\cal T}(r)$ depends upon the spin weight of the
radiation.  It is constructed by projecting the radiation source onto
legs of the Newman-Penrose null tetrad, ${\bf l}$, ${\bf n}$, ${\bf
m}$, and ${\bf\bar m}$.  A useful representation of the tetrad in
Boyer-Lindquist coordinates is {\cite{chandra_mtbh}}
\begin{eqnarray}
l_\alpha &=& \left[1,-{\Sigma\over\Delta},0,-a\sin^2\theta\right]\;,
\nonumber\\
n_\alpha &=& {1\over2}\left[{\Delta\over\Sigma},1,0,
-{a\Delta\sin^2\theta\over\Sigma}\right]\;,
\nonumber\\
m_\alpha &=& {1\over\sqrt{2}(r + i a \cos\theta)}
\left[ia\sin\theta,0,-\Sigma,-i(r^2 + a^2)\sin\theta\right]\;.
\label{eq:NPtetrad}
\end{eqnarray}
The tetrad legs ${\bf l}$ and ${\bf n}$ represent ingoing and outgoing
null vectors, respectively.  Quantities constructed by projecting onto
${\bf l}$ correspond to ingoing radiation and their sources; they map
to positive $s$.  Likewise, projection onto ${\bf n}$ corresponds to
outgoing radiation and their sources\footnote{Because one can
transform between positive and negative $s$ solutions, one can
actually develop both ingoing and and outgoing radiation with a single
source term.}, and map to negative $s$.  See Ref.\ {\cite{teuk73}} for
details.

Because Eq.\ (\ref{eq:teuk_eqn_inhom}) is in self-adjoint form, one
can construct its solution by the method of Green's functions
{\cite{arfken}}.  This means that one needs to know only the solutions
to the homogeneous equation,
\begin{equation}
\Delta^{-s}\left(\Delta^{s+1} R'\right)' - V_T(r) R = 0\;,
\label{eq:teuk_eqn}
\end{equation}
in addition to the source.  One does this by adapting the generic
solution, given in Eqs.\ (\ref{eq:teuk_soln_inf}) and
(\ref{eq:teuk_soln_horiz}), to the appropriate boundary conditions: no
radiation may come in from infinity and none may come out from the
event horizon.  In other words, there exist two solutions, $R^H(r)$
and $R^\infty(r)$, whose asymptotic forms are
\begin{eqnarray}
R^H(r) &=& B^{\rm hole} \Delta^{-s} e^{-i p r^*}\;,\qquad
r \to r_+\nonumber\\
&=& B^{\rm out} {e^{i\omega r^*}\over r^{2s+1}} + B^{\rm in}
{e^{-i \omega r^*}\over r}\;,\qquad
r \to \infty\;,\label{eq:RHdef}\\
R^\infty(r)&=& D^{\rm out} e^{ip r^*} + D^{\rm in}\Delta^{-s}
e^{-i p r^*}\;,\qquad
r \to r_+ \nonumber\\
&=& D^\infty {e^{i \omega r^*}\over r^{2s + 1}}\;,\qquad r \to\infty\;.
\label{eq:Rinfdef}
\end{eqnarray}
The solution to the inhomogeneous equation (\ref{eq:teuk_eqn_inhom})
which one constructs from Eqs.\ (\ref{eq:RHdef}), (\ref{eq:Rinfdef}),
and the source ${\cal T}(r)$ is conveniently written
\begin{equation}
R(r) = Z^H(r) R^\infty(r) + Z^\infty(r)R^H(r)\;,
\label{eq:R_general_soln}
\end{equation}
where
\begin{eqnarray}
Z^H(r) &=& {1\over 2i\omega B^{\rm in} D^\infty}
\int_{r_+}^r dr'\,\Delta(r')^sR^H(r'){\cal T}(r')\;,
\nonumber\\
Z^\infty(r) &=& {1\over 2i\omega B^{\rm in} D^\infty}
\int_r^\infty dr'\,\Delta(r')^sR^\infty(r'){\cal T}(r')\;.
\end{eqnarray}

Using Eq.\ (\ref{eq:R_general_soln}) one can construct ${_s}\Psi$.
This quantity is related to a radiation field of spin weight $s$; the
details of that relation depend upon the value of $s$.  Typically,
${_s}\Psi$ is constructed by projecting a tensor describing the
radiation onto legs of the Newman-Penrose null tetrad.  For example,
${_0}\Psi = \Phi$, a massless scalar field.  No projections are needed
in this case.  For $s = \pm 1$, we have
\begin{eqnarray}
{_1}\Psi &=& \phi_0 = F_{\mu\nu}l^\mu m^\nu\;,
\nonumber\\
{_{-1}}\Psi &=& (r - i a \cos\theta)^2\phi_2
= (r - i a \cos\theta)^2F_{\mu\nu}n^\mu {\bar m}^\nu\;,
\label{eq:em_fields}
\end{eqnarray}
where $F_{\mu\nu}$ is the electromagnetic field tensor.  [There is a
third projection, $\phi_1 = (1/2)F_{\mu\nu}(l^\mu n^\nu + {\bar m}^\mu
m^\nu)$.  It does not describe the radiative degrees of freedom of the
electromagnetic field, and so is of less interest here.]  For $s = \pm
2$, the radiative quantities are
\begin{eqnarray}
{_2}\Psi &=& \psi_0 = -C_{\alpha\beta\gamma\delta}
l^\alpha m^\beta l^\gamma m^\delta\;,
\nonumber\\
{_{-2}}\Psi &=& (r - i a \cos\theta)^4\psi_4
= -(r - i a \cos\theta)^4C_{\alpha\beta\gamma\delta}
n^\alpha {\bar m}^\beta n^\gamma {\bar m}^\delta\;.
\label{eq:grav_fields}
\end{eqnarray}
The tensor $C_{\alpha\beta\gamma\delta}$ is the Weyl component of the
spacetime's curvature.  The quantities $\psi_i$, with $i$ an integer
from 0 to 4, are the Newman-Penrose projections of the Weyl curvature
(see Ref.\ {\cite{chandra_mtbh}}).  For unperturbed black hole
spacetimes, all components except $\psi_2 =
-C_{\alpha\beta\gamma\delta}l^\alpha m^\beta {\bar m}^\gamma n^\delta
= M/(r - i a \cos\theta)^3$ can be set to zero with an appropriate
choice of gauge.  This is the non-radiative ``background'' component
of the curvature; the perturbations $\psi_0$ and $\psi_4$ represent
radiation on the background.

The solution for the (linear) evolution of radiation of spin weight
$s$ in a Kerr black hole spacetime is thus completely described by
construction of the source ${\cal T}(r)$ appropriate to that spin
weight and construction of the homogeneous solutions $R^H(r)$ and
$R^\infty(r)$.  As discussed in the Introduction --- and, as should be
clear from the asymptotic solutions (\ref{eq:RHdef}) and
(\ref{eq:Rinfdef}) --- it is very difficult to build these solutions
in a numerical integration.  The remainder of this paper is devoted to
methods for constructing $R^H(r)$ and $R^\infty(r)$ by finding
transformations that relate the Teukolsky solution $R(r)$ to solutions
of equations with short-ranged potentials.

\section{Results for Schwarzschild holes}
\label{sec:schwarzschild}

The Teukolsky equation for Schwarzschild black holes is
\begin{equation}
\Delta^{-s} \left(\Delta^{s+1} R'\right)' - V_{TS}(r) R = 0\;,
\label{eq:teukeqn_ss}
\end{equation}
where
\begin{equation}
V_{TS}(r) = \lambda - 4isr\omega + [2is(r - M)\omega -
(r\omega)^2]/f\;,
\label{eq:teukpot_ss}
\end{equation}
and $\lambda = \lambda(a = 0) = l(l+1) - s(s+1)$.

We would like to find rules that allow us to obtain $R$ given a
solution $X$ of the Regge-Wheeler equation, (\ref{eq:reggewheeler}).
To do so, first define the quantity
\begin{equation}
\chi \equiv {X\over r\sqrt{\Delta^s}}\;.
\label{def:chi_ss}
\end{equation}
If $X$ satisfies the Regge-Wheeler equation, it is straightforward to
show that $\chi$ satisfies
\begin{equation}
\Delta^{-s} \left(\Delta^{s+1}\chi'\right)' - U_{\chi S}(r) \chi = 0\,
\label{eq:chieqn_ss}
\end{equation}
where
\begin{equation}
U_{\chi S}(r) = \lambda + {1\over f}\left[s^2\left({3 M^2\over r^2} -
{2 M\over r}\right) - (r\omega)^2\right]\;.
\label{eq:chipot_ss}
\end{equation}

By direct substitution, one can show that $\chi$ can be transformed to
$R$, and vice versa, via
\begin{mathletters}
\begin{eqnarray}
{s < 0:}\qquad
\chi &=& \left(r\sqrt{\Delta}\right)^{|s|}{\cal D}^{|s|}_-
\left[{R\over r^{|s|}}\right]\;,\nonumber\\
R &=& \left({\Delta\over r}\right)^{|s|}{\cal D}^{|s|}_+\left[
\left({r\over\sqrt{\Delta}}\right)^{|s|}\chi\right]\;,
\label{eq:sstrans_sneg}\\
{s = 0:}\qquad
\chi &=& R\;,
\label{eq:sstrans_szero}\\
{s > 0:}\qquad
\chi &=& \left({r\over\sqrt{\Delta}}\right)^s{\cal D}^s_+
\left[\left({\Delta\over r}\right)^sR\right]\;,\nonumber\\
R &=& \left({1\over r}\right)^s{\cal D}^s_-
\left[\left(r\sqrt{\Delta}\right)^s\chi\right]\;,
\label{eq:sstrans_spos}
\end{eqnarray}
\end{mathletters}
where
\begin{equation}
{\cal D}_{\pm} = d/dr \pm i\omega/f.
\label{eq:ss_dee}
\end{equation}
For $s = -1$ and $s = -2$, Eq.\ (\ref{eq:sstrans_sneg}) reduces to the
Chandrasekhar transformation (see Ref.\ {\cite{poisson93}} for $s =
-2$, Ref.\ {\cite{leonard_poisson}} for $s = -1$).  Equations
(\ref{eq:sstrans_sneg}) -- (\ref{eq:sstrans_spos}) serve as guidelines
that will be used to fix the form of the transformation rules for Kerr
black holes.  Note that the transformations from $R$ to $\chi$ can be
written
\begin{equation}
\chi = \alpha R + \beta\Delta^{s+1} R'
\label{eq:generic_trans_rule}
\end{equation}
by repeatedly using Eq.\ (\ref{eq:teukeqn_ss}) to eliminate
derivatives of second order and higher.  The resulting functions
$\alpha$ and $\beta$ may become rather complicated, particularly for
large values of $|s|$, but the general operation is straightforward.
(The factor $\Delta^{s+1}$ is inserted for later convenience.)

\section{Perturbation equation for Kerr holes}
\label{sec:kerr}

Guided by Eq.\ (\ref{eq:generic_trans_rule}), let us assume that
functions $\alpha$ and $\beta$ can be found that transform the {\it
Kerr} solution $R$ to solutions $\chi$ of some other equation.  By
generalizing the relation (\ref{def:chi_ss}) to a form appropriate for
the Kerr metric and rewriting all derivatives in terms of $r^*$ we
will come to an equation with short-ranged potentials governing the
behavior of a function $X(r)$.  This is the generalized
Sasaki-Nakamura (GSN) equation.  It will depend explicitly on the
(currently unspecified) functions $\alpha$ and $\beta$.  These
functions will be specified by requiring that the transformation rule
satisfy a form which reduces to Eqs.\ (\ref{eq:sstrans_sneg}) --
(\ref{eq:sstrans_spos}) when $a = 0$.  This guarantees that solutions
to the GSN equation are equivalent to solutions of the Regge-Wheeler
equation in the Schwarzschild limit.

To begin, differentiate $\chi$ and use Eq.\ (\ref{eq:teuk_eqn}) to
eliminate the second derivative of $R$.  The resulting equations for
$\chi$ and $\chi'$ can be gathered neatly into matrix form:
\begin{equation}
\left(\matrix{\chi \cr \chi'}\right) =
\left(\matrix{
\alpha & \beta\Delta^{s+1}\cr
\alpha' + \beta V_T \Delta^s & \alpha + \beta' \Delta^{s+1}\cr
}\right)
\left(\matrix{R \cr R'}\right)\;.
\label{eq:R_to_chi}
\end{equation}
A nice feature of Eq.\ (\ref{eq:R_to_chi}) is that the inverse
solution is rather obvious:
\begin{equation}
\left(\matrix{R \cr R'}\right) =
{1\over\eta}\left(\matrix{
\alpha + \beta' \Delta^{s+1} & -\beta\Delta^{s+1} \cr
-(\alpha' + \beta V_T \Delta^s) & \alpha \cr
}\right)
\left(\matrix{\chi \cr \chi'}\right)\;,
\label{eq:chi_to_R}
\end{equation}
where
\begin{equation}
\eta = \alpha\left(\alpha + \beta'\Delta^{s+1}\right)
-\beta\Delta^{s+1}\left(\alpha' + \beta V_T \Delta^s\right)
\label{eq:eta_def}
\end{equation}
is the determinant of the matrix in Eq.\ (\ref{eq:R_to_chi}).

Differentiating again and massaging the resultant expression gives us
a second-order differential equation for $\chi$:
\begin{equation}
\Delta^{-s}\left(\Delta^{s+1}\chi'\right)'
- \Delta {_{s}}F_1(r) \chi' - {_{s}}U_1(r)\chi = 0\;.
\label{eq:chieqn_kerr}
\end{equation}
The potentials ${_{s}}F_1(r)$ and ${_{s}}U_1(r)$ are given by
\begin{eqnarray}
{_{s}}F_1(r) &=& \eta'/\eta\;,\nonumber\\
{_{s}}U_1(r) &=& V_T + {1\over\beta\Delta^2}
\left[\left(2\alpha + \beta'\Delta^{s+1}\right)' - {\eta'\over\eta}
\left(\alpha + \beta'\Delta^{s+1}\right)\right]\;.
\label{eq:F1_and_U1}
\end{eqnarray}

Next, generalize Eq.\ (\ref{def:chi_ss}) to the Kerr form
\begin{equation}
\chi \equiv {X\over{\sqrt{(r^2 + a^2)\Delta^s}}}\;.
\label{def:chi_kerr}
\end{equation}
Using this to replace $\chi$ for $X$ in Eq.\ (\ref{eq:chieqn_kerr})
and then replacing derivatives in $r$ with derivatives in $r^*$ with
the rule
\begin{equation}
{d\over dr} = {(r^2 + a^2)\over\Delta}{d\over dr^*}
\label{eq:r_to_rstar}
\end{equation}
yields the GSN equation:
\begin{equation}
{d^2X\over dr^{*2}} - {_{s}}F(r) {dX\over dr^*} - {_{s}}U(r) X = 0\;.
\label{eq:gsn_eqn}
\end{equation}
The potentials are
\begin{eqnarray}
{_{s}}F(r) &=& {\Delta {_{s}}F_1(r)\over r^2 + a^2}\;,\nonumber\\
{_{s}}U(r) &=& {\Delta {_{s}}U_1(r)\over(r^2 + a^2)^2} + {_{s}}G(r)^2 +
{\Delta d{_{s}G}/dr\over r^2 + a^2} -
{\Delta{_{s}G(r)}{_{s}F_1(r)}\over r^2 + a^2}\;.
\label{eq:gsn_potentials}
\end{eqnarray}
The function ${_{s}}G(r)$ is
\begin{equation}
{_{s}}G(r) = {r\Delta\over(r^2 + a^2)^2} + {s(r - M)\over r^2 + a^2}\;;
\label{eq:G_function}
\end{equation}
the functions ${_{s}}F_1(r)$ and ${_{s}}U_1(r)$ are from Eq.\
(\ref{eq:F1_and_U1}).  When $s = -2$, all functions reduce to those
given by Sasaki and Nakamura (see Ref.\ {\cite{sasaknak}}).

All of the quantities which have been derived to this point depend
upon the as-yet-undetermined functions $\alpha$ and $\beta$.  We fix
these functions by requiring that they affect a transformation between
$R$ and $\chi$ which, as $a \to 0$, reduces to Eqs.\
(\ref{eq:sstrans_sneg}) -- (\ref{eq:sstrans_spos}).  A useful
generalization of these transformations is
\begin{mathletters}
\begin{eqnarray}
{s < 0:}\qquad
\chi &=& \left(\sqrt{(r^2 + a^2)\Delta}\right)^{|s|}g_0(r)J_-
\left[g_1(r)J_-\left[g_2(r) \ldots J_-\left[{g_{|s|}(r)R
\over\left(\sqrt{r^2 + a^2}\right)^{|s|}}\right]\right]\right]\;,
\label{eq:kerrtrans_sneg}\\
{s = 0:}\qquad
\chi &=& g_0(r) R\;,
\label{eq:kerrtrans_szero}\\
{s > 0:}\qquad
\chi &=& \left(\sqrt{r^2 + a^2\over\Delta}\right)^s g_0(r)
J_+\left[g_1(r)J_+\left[g_2(r) \ldots J_+\left[g_s(r)
\left({\Delta\over\sqrt{r^2 + a^2}}\right)^s R
\right]\right]\right]\;,
\label{eq:kerrtrans_spos}
\end{eqnarray}
\end{mathletters}
where the operator
\begin{equation}
J_\pm = d/dr \pm i K(r)/\Delta
\label{eq:kerr_kay}
\end{equation}
generalizes ${\cal D}_\pm$ to Kerr.  The $s = +2$ transformation rule,
for example, is
\begin{equation}
\chi = g_0(r) {(r^2 + a^2)\over\Delta} J_+\left[g_1(r) J_+\left[
g_2(r){\Delta^2 \over r^2 + a^2}R\right]\right]\;;
\label{eq:example_s=2}
\end{equation}
an example for $s = -1$ is given in Sec.\ {\ref{sec:electromagnetic}}.

To now specify $\alpha$ and $\beta$, one must pick functions $g_i(r)$
and then repeatedly use Eq.\ (\ref{eq:teuk_eqn}) to eliminate
derivatives of second order and higher in Eqs.\
(\ref{eq:kerrtrans_sneg}) -- (\ref{eq:kerrtrans_spos}).  The resultant
expressions for $\alpha$ and $\beta$ will be, in general, quite
complicated; examples are discussed in Secs.\ {\ref{sec:scalar}} and
{\ref{sec:electromagnetic}}.  The functions $g_i(r)$ must be chosen so
that they become constant in the Schwarzschild limit, and lead to
potentials $F(r)$ and $U(r)$ which are short-ranged [{\it i.e.}, fall
off at a rate $O(1/r^2)$ or faster as $r \to \infty$].  In practice,
choosing $g_i(r) = 1$ or $g_i(r) = (r^2 + a^2)/r^2$ appears to lead to
well-behaved potentials; some experimentation may be needed to make
useful choices.

\section{Scalar radiation}
\label{sec:scalar}

For scalar radiation, $s = 0$, the functions $R$, $X$, and $\chi$ have
the following relationship:
\begin{equation}
g_0(r) R = \chi = {X\over\sqrt{r^2 + a^2}}\;.
\label{eq:scalarfuncs}
\end{equation}
A good choice is $g_0(r) = 1$.  The functions $\alpha$ and $\beta$
[cf.\ Eq.\ (\ref{eq:generic_trans_rule})] are then given by
\begin{equation}
\alpha = 1\;,\qquad
\beta = 0\;.
\label{eq:scalar_alpha_beta}
\end{equation}
From this, it follows that
\begin{mathletters}
\begin{eqnarray}
\eta &=& 1\;,
\label{eq:scalar_eta}\\
{_0}F_1 &=& 0\;,
\label{eq:scalar_F}\\
{_0}U_1 &=& V_T\;,
\label{eq:scalar_U1}\\
{_0}G &=& {r\Delta\over r^2 + a^2}\;.
\label{eq:scalar_G}
\end{eqnarray}
\end{mathletters}
The potentials ${_0}F(r)$ and ${_0}U(r)$ are given by substituting
Eqs.\ (\ref{eq:scalar_eta}) -- (\ref{eq:scalar_G}) into Eq.\
(\ref{eq:gsn_potentials}).  For large $r$,
\begin{equation}
{_0}U(r) = -\omega^2 + {\lambda + 2 a m \omega\over r^2} +
{2M\left(1 - \lambda\right)\over r^3} + O(1/r^4)\;;
\label{eq:scalar_U_large_r}
\end{equation}
clearly, ${_0}F(r) = 0$ for all $r$.  Hence, the potentials are
short-ranged.  When $a = 0$, ${_0}U(r)$ reduces to $-\omega^2 + V_{\rm
RW}(r,s = 0)$.  Thus, it reduces to the Regge-Wheeler equation in the
Schwarzschild limit, as it was supposed to.

The asymptotic solutions to the $s = 0$ GSN equation are simple plane
waves:
\begin{eqnarray}
X^H(r) &=& e^{-ipr^*}\;,\qquad r \to r_+\;,
\nonumber\\
&=& A^{\rm out} {\bar P}_0(r) e^{i\omega r^*} +
A^{\rm in} P_0(r) e^{-i\omega r^*}\;,\qquad r \to \infty\;;
\label{eq:scalar_XH}\\
X^\infty(r) &=& C^{\rm out}e^{ipr^*} + C^{\rm in}e^{-i p r^*}\;,
\qquad r \to r_+\;,\nonumber\\
&=& {\bar P}_0(r) e^{i\omega r^*}\;,\qquad r \to \infty\;.
\label{eq:scalar_Xinf}
\end{eqnarray}
The function
\begin{equation}
P_0(r) = 1 + {{\cal A}_0\over\omega r} + {{\cal B}_0\over(\omega r)^2} +
{{\cal C}_0\over(\omega r)^3} + \ldots
\label{eq:scalar_P_def}
\end{equation}
allows us to more accurately describe the behavior of $X^{H,\infty}$
near infinity.  This is useful both to improve numerical computations
and to derive certain relations between the amplitudes of the
Teukolsky solution and the GSN solution.  The first three coefficients
have the values
\begin{eqnarray}
{\cal A}_0 &=& -{i\over2}\left(\lambda + 2am\omega\right)\;,
\nonumber\\
{\cal B}_0 &=& -{1\over8}\left[\lambda^2 - \lambda\left(2 -
4am\omega\right) - 4\left[am\omega - i M \omega -
am\omega\left(am\omega + 2 i M \omega\right)\right]\right]\;,
\nonumber\\
{\cal C}_0 &=& -{i\over6}\left[{\cal B}_0\left(\lambda - 6 + 2am\omega
+ 8iM\omega\right) - 4\left(M\omega\right)^2 - 2{\cal
A}_0M\omega\left(\lambda - 6\right) \right.\nonumber\\
& &\left. - \left(a\omega\right)^2\left(\lambda - 1 + m^2 +
2am\omega\right)\right]\;.
\nonumber\\
\label{eq:scalar_ABC}
\end{eqnarray}

\section{Electromagnetic radiation}
\label{sec:electromagnetic}

For electromagnetic radiation, $s = -1$, the functions $R$, $X$, and
$\chi$ exhibit the following relationships:
\begin{eqnarray}
\chi &=& \sqrt{\Delta\over r^2 + a^2} X\;,
\label{eq:em_chi_X}\\
\chi &=& \alpha R + \beta R'
\label{eq:em_chi_R_1}\\
&=& g_0(r)\sqrt{(r^2 + a^2)\Delta}J_-\left[{g_1(r) R\over\sqrt{r^2 +
a^2}}\right]\;.
\label{eq:em_chi_r_2}
\end{eqnarray}
From this, we can read off
\begin{eqnarray}
\alpha &=& g_0\sqrt{\Delta}\left[g_1' - {r g_1\over r^2 + a^2} -
{i g_1 K\over\Delta}\right]\;,
\nonumber\\
\beta &=& g_0 g_1\sqrt{\Delta}\;.
\label{eq:em_alpha_beta}
\end{eqnarray}
A useful choice for $g_0$ and $g_1$ is
\begin{equation}
g_0(r) = {(r^2 + a^2)\over r^2}\;,\qquad
g_1(r) = 1\;.
\label{eq:em_gs}
\end{equation}
The function $\eta$ that follows from these choices is
\begin{equation}
\eta = c_0 + c_1/r + c_2/r^2 + c_3/r^3 + c_4/r^4\;,
\label{eq:em_eta}
\end{equation}
where
\begin{eqnarray}
c_0 &=& -\lambda\;,\nonumber\\
c_1 &=& -2iam\;,\nonumber\\
c_2 &=& a^2(1 - 2\lambda)\;,\nonumber\\
c_3 &=& -2a^2(M + i a m)\;,\nonumber\\
c_4 &=& a^4(1 - \lambda)\;.
\label{eq:em_eta_cofs}
\end{eqnarray}

Using this $\eta$ and
\begin{equation}
{_{-1}}G(r) = {r\Delta\over(r^2 + a^2)^2} - {(r - M)\over r^2 + a^2}\;,
\label{eq:em_G}
\end{equation}
it is straightforward to construct the functions ${_{-1}}F_1(r)$,
${_{-1}}U_1(r)$, ${_{-1}}F(r)$, and ${_{-1}}U(r)$.  The results are
rather complicated and are not given here.  When $r$ is large,
\begin{eqnarray}
{_{-1}}F(r) &=& -{2 i a m\over\lambda r^2} + {2a\left[2 i m M \lambda
- a\left(2 m^2 + 2\lambda^2 - \lambda\right)\right]\over\lambda^2r^3}
+ O(1/r^3)\;,
\nonumber\\
{_{-1}}U(r) &=& -\omega^2 + {\lambda^2 + 2am\omega(\lambda +
1)\over\lambda r^2} - {2\left[i a^2(2 m^2 - \lambda) + M(\lambda^3 +
2am\omega\lambda)\right]\over\lambda^2r^3} + O(1/r^4)\;.
\nonumber\\
\label{eq:em_F_U}
\end{eqnarray}
Both ${_{-1}}F(r)$ and ${_{-1}}U(r)$ are short ranged.  When $a = 0$,
${_{-1}}F(r) = 0$ and ${_{-1}}U(r) = -\omega^2 + V_{RW}(r, s = -1)$.

The solutions $X^{H,\infty}$ are, in the limits $r \to r_+$ and $r \to
\infty$, essentially identical to those given in Eqs.\
(\ref{eq:scalar_XH}) and (\ref{eq:scalar_Xinf}); one need only change
the subscript on the $P$ function to $-1$.  The corresponding
coefficients in $P_{-1}(r)$ are
\begin{eqnarray}
{\cal A}_{-1} &=& -{i\over2}\left(\lambda + 2am\omega\right)\;,
\nonumber\\
{\cal B}_{-1} &=& -{1\over8}\left[\lambda^2 - \lambda\left(2 -
4am\omega\right) - 4a\omega\left[m - 2 i m M \omega + a\omega \left(2 -
m^2\right)\right]\right]\;,
\nonumber\\
{\cal C}_{-1} &=& -{i\over6}\left[{\cal B}_{-1}\left(\lambda - 6 +
2am\omega + 8iM\omega\right) + 2{\cal A}_{-1}\left[M\omega\left(5 -
\lambda\right) + ia\omega\left(2 a\omega +
m/\lambda\right)\right]\right.
\nonumber\\
& &\left. - (a\omega)^2\left[\lambda - 5 + 8 i M \omega + 2 a m \omega
+ m^2(\lambda + 2)/\lambda\right]\right]\;.
\label{eq:em_ABC}
\end{eqnarray}

\section{Conclusion}
\label{sec:conclusion}

In this paper, I have shown how the Sasaki-Nakamura short-ranged
equation for gravitational perturbations to a rotating black hole may
be generalized to arbitrary spin weight radiation.  Of course, there
is no particular physical motivation for choosing radiation spin of
magnitude greater than $2$; this approach is taken simply so that one
can write down a single rule which encompasses all physically
interesting radiation fields, much as the Teukolsky equation itself
encompasses all spin weights.  Efficient numerical computation of
Teukolsky equation solutions now reduces to a simple recipe.  First,
following the analysis in Sec.\ {\ref{sec:kerr}}, develop the
potentials needed in the GSN equation, Eq.\ (\ref{eq:gsn_eqn}).
Examples are given for $s = 0$ and $s = -1$.  Integrate Eq.\
(\ref{eq:gsn_eqn}) for the GSN solution $X$.  Transform to the
variable $\chi$ using Eq.\ (\ref{def:chi_kerr}).  Then construct the
Teukolsky solution $R$ using Eq.\ (\ref{eq:R_to_chi}).

One application of these results may be to extend the mode sum
regularization scheme described in Ref.\ {\cite{barack_ori}} to self
forces computed in Kerr spacetimes.  Calculations that employ scalar
or electromagnetic charges and fields are generally simpler than the
gravitational self force calculations, which are of great interest for
researchers studying gravitational-wave sources.  The electromagnetic
perturbation equation given in Sec.\ {\ref{sec:electromagnetic}} may
be of astrophysical interest, particularly when coupled to an
appropriate source.

\acknowledgements

I thank Lior Burko for encouraging me to find transformation rules and
perturbation equations for generic $s$, Eric Poisson for valuable
comments, and Manuela Campanelli for pointing me to some useful
references.  I am also very grateful to Steven Detweiler for valuable
comments on the history of black hole perturbation studies which I
have used to correct some of the discussion in the Introduction.  The
package {\sc Mathematica} was used to aid some of the calculations.
This research was supported by NSF Grant AST-9731698 and NASA Grants
NAG5-7034 and NAGW-4268.


\begin{references}

\bibitem{teuk73} S.\ A.\ Teukolsky, Astrophys.\ J.\ {\bf 185}, 635 (1973).

\bibitem{pullin} J.\ Pullin, Prog.\ Theor.\ Phys.\ Suppl.\ {\bf 136},
	p.\ 107 (1999); also gr-qc/9909021.

\bibitem{spherical1} W.\ B.\ Campbell and T.\ Morgan, Physica {\bf 53},
	264 (1971).

\bibitem{spherical2} J.\ N.\ Goldberg {\it et al.}, J.\ Math.\ Phys.\
	{\bf 8}, 2155 (1967).

\bibitem{sah_kerr_gwI} S.\ A.\ Hughes, Phys.\ Rev.\ D, in press;
	also gr-qc/9910091.

\bibitem{leonard_poisson} S.\ W.\ Leonard and E.\ Poisson, Phys.\ Rev.\ D
	{\bf 56}, 4789 (1997).

\bibitem{teukpress} S.\ A.\ Teukolsky and W.\ H.\ Press, Ap.\ J.\ {\bf 193},
	443 (1974).

\bibitem{pressteuk} W.\ H.\ Press and S.\ A.\ Teukolsky, Ap.\ J.\ {\bf 185},
	649 (1973).

\bibitem{chandra_transform} S.\ Chandrasekhar, Proc.\ R.\ Soc.\ London
	{\bf A343}, 289 (1975).

\bibitem{leaver85} E.\ W.\ Leaver, Proc.\ Roy.\ Soc.\ London {\bf A402},
	285 (1985).

\bibitem{poisson93} E.\ Poisson, Phys.\ Rev.\ D {\bf 47}, 1497 (1993).

\bibitem{sasaknak} M.\ Sasaki and T.\ Nakamura, Phys.\ Lett.\ {\bf 89A},
	68 (1982); M.\ Sasaki and T.\ Nakamura, Prog.\ Theor.\ Phys.\
	{\bf 67}, 1788 (1982).

\bibitem{chandra_det76} S. Chandrasekhar and S.\ L.\ Detweiler, Proc.\
	Roy.\ Soc.\ London {\bf A350}, 165 (1976).

\bibitem{det77} S.\ L.\ Detweiler, Proc.\ Roy.\ Soc.\ London {\bf A352},
	381 (1977).

\bibitem{chandra_cent} S.\ Chandrasekhar, in {\it General Relativity
	--- An Einstein Centenary Survey}, edited by S.\ W.\ Hawking
	and W.\ Israel (Cambridge, England, 1979), chap.\ 7.

\bibitem{chandra_mtbh} Much of the work by Chandrasekhar and
	collaborators is surveyed in S.\ Chandrasekhar, {\it The
	Mathematical Theory  of Black Holes} (Oxford University Press,
	New York, 1983), chaps.\ 8 -- 9.

\bibitem{det_pc} S.\ Detweiler, private communication.

\bibitem{camp_lousto} M.\ Campanelli and C.\ O.\ Lousto, Phys.\ Rev.\ D
	{\bf 56}, 6363 (1997).

\bibitem{kojima_nakamura} Y.\ Kojima and T.\ Nakamura, Phys.\ Lett.\
	{\bf 96A}, 335 (1983); Y.\ Kojima and T.\ Kakamura, Prog.\
	Theor.\ Phys.\ {\bf 71}, 79 (1984).

\bibitem{shibata94} M.\ Shibata, Phys.\ Rev.\ D {\bf 50}, 6297 (1994).

\bibitem{msstt97} Y.\ Mino {\it et al.}, Prog.\ Theor.\ Phys.\ Supplement
	No.\ 128 (1997), chap.\ 1.

\bibitem{kennefick} D.\ Kennefick, Phys.\ Rev.\ D {\bf 58}, 4012 (1998);
	also gr-qc/9805102.

\bibitem{mst97} Y.\ Mino, M.\ Sasaki, and T.\ Tanaka, Phys.\ Rev.\ D
	{\bf 55}, 3457 (1997).

\bibitem{quinn_wald} T.\ C.\ Quinn and R.\ M.\ Wald, Phys.\ Rev.\ D
	{\bf 56}, 3381 (1997).

\bibitem{and_flan_ott} W.\ G.\ Anderson, E.\ E.\ Flanagan, and
	A.\ C.\ Ottewill, in preparation.

\bibitem{wiseman} A.\ G.\ Wiseman, Phys.\ Rev.\ D, in press; also
	gr-qc/0001025.

\bibitem{ori} A.\ Ori, Phys.\ Lett.\ A {\bf 202}, 347 (1995); A.\
	Ori, Phys.\ Rev.\ D {\bf 55}, 3444 (1997).

\bibitem{barack_ori} L.\ Barack and A.\ Ori, Phys.\ Rev.\ D, in press;
	also gr-qc/9912010.

\bibitem{burko1} L.\ M.\ Burko, Class.\ Quantum Grav. {\bf 17}, 227 (2000).

\bibitem{burko2} L.\ M.\ Burko, in preparation; see also L.\ M.\ Burko,
	gr-qc/9911089.

\bibitem{arfken} G.\ Arfken, {\it Mathematical Methods for Physicists}
	(Academic Press, Orlando, 1985), chap.\ 16.

\end{references}
\end{document}